\documentclass[graybox]{svmult}

\usepackage{type1cm}        
\usepackage{makeidx}       
\usepackage{graphicx}      
		 
\usepackage{multicol}        
\usepackage[bottom]{footmisc}

\usepackage{newtxtext}       %
\usepackage[varvw]{newtxmath}       

\usepackage{cite}       

\usepackage{subfiles}

\makeindex     

\newcommand{\kon}{k_\textrm{on}}
\newcommand{\koff}{k_\textrm{off}}

\newcommand{\kopti}{k_\textrm{off}^*}

\newcommand{\cDNA} {c_\textrm{DNA}}
\newcommand{\cbulk} {c_\textrm{bulk}}
\newcommand{\ctarget} {c_\textrm{target}}

\newcommand{\cDNAbar} {\bar c_\textrm{DNA}}
\newcommand{\cbulkbar} {\bar c_\textrm{bulk}}
\newcommand{\ctargetbar} {\bar c_\textrm{target}}
\newcommand{\konbar}{\bar k_\textrm{on}}

\newcommand{\cAntenna}{\bar c (x)}

\newcommand{\be}{\begin{equation}}
\newcommand{\ee}{\end{equation}}

\begin{document}

\title*{
	Exploring the benefits of DNA-target search with antenna
	}
\author{
	Lucas Hedstr\"om and Ludvig Lizana (\today)
	}
\institute{
	Integrated Science Lab, Department of Physics, 
	Ume\aa~University, SE-901 87 Ume\aa, Sweden,
	\email{ludvig.lizana@umu.se}
	}
%
%
\maketitle

\begin{abstract}
~The most common gene regulation mechanism is when a protein binds to a regulatory sequence to change RNA transcription. However, these sequences are short relative to the genome length, so finding them poses a challenging search problem. This chapter presents two mathematical frameworks capturing different aspects of this problem. First, we study the interplay between diffusional flux through a target where the searching proteins get sequestered on DNA far from the target because of non-specific interactions. From this model, we derive a simple formula for the optimal protein-DNA unbinding rate, maximizing the particle flux. Second, we study how the flux flows through a target on a single antenna with variable length. Here, we identify a non-trivial logarithmic correction to the linear behavior relative to the target size proposed by Smoluchowski's flux formula.
\end{abstract}

\section{Introduction}
\label{sec:introduction}
Cells use designated proteins called transcription factors (TFs) to regulate gene expression.   These TFs adjust the expression levels of genes in response to external and internal triggers such as changed nutrient or toxin concentrations, viral attacks, the physical presence of other cells, or damaged DNA.  Mechanistically, TFs regulate genes by binding to short operator sequences typically located a few base pairs downstream of the gene's transcription start site from where they may hinder or help RNA polymerases to transcribe the gene into messenger RNA.  However, because DNA is so long,  TFs face a considerable challenge in finding these operator sites.  And in some cases, they must find them quickly to counteract critical changes. Also, once found, they must stay bound long enough to influence RNA transcription (or return sufficiently often). 

For illustrative purposes, let's discuss the TF search problem in \textit{Escherichia coli} bacteria (E. coli) using actual numbers. E. coli encodes their environments in concentrations of about 300 TF types that regulate the expression levels of approximately 4,200 genes \cite{milo2015cell}. Since copy numbers typically are low,  $\sim 10^2$ per TF species (geometric average), and operator sites are scattered across a $5 \cdot 10^6$ basepair (bp) long DNA, each TF must scan about 50,000 bps per second to achieve second-scale response times. Translated into a one-dimensional diffusion constant, this corresponds to
$D_\mathrm{1D} \sim 5\cdot10^4 \mathrm{bp}^2 \mathrm {sec}^{-1}$,
which is not too far from with \textit{in vitro} single-particle tracking measurements on the LacI repressor yielding 
$D_\mathrm{1D}^\textrm{LacI} = (4.0 \pm 0.9)\cdot10^5 \mathrm{bp}^2 \mathrm {sec}^{-1}$\cite{elf2007probing}. 
However,  this value appears to be about ten times higher \textit{in vivo} \cite{elf2007probing}, indicating that linear diffusion cannot be the primary search mechanism.  The standard explanation for this discrepancy is that LacI diffuses along short DNA segments and occasionally transfers between segments through the cytoplasm. Since the apparent diffusion constant represents a weighted average between two diffusion mechanisms, one slow (DNA) and one fast (cytoplasm), it should exceed $D_\mathrm{1D}^\textrm{LacI}$.

Next, to estimate realistic binding constants, we note that $10^2$ proteins correspond to approximately $c^\mathrm{TF} = 0.1$ micromolar ($\mu \mathrm M$) intracellular concentration (1 protein in E. coli's volume is 1.6 $10^{-3}\mu \mathrm M$ \cite{phillips2012physical}). This suggests that  TF-operator binding constants must be $K_\textrm{op}^\textrm{TF}\sim 0.01 \mu \mathrm M$  to ensure a 90\%  occupation probability%
\footnote{Assuming first-order binding kinetics, the binding probability is $p_b=c_\textrm{TF}/(c_\textrm{TF}+K_\textrm{op}^\textrm{TF})$, or $K_\textrm{op}^\textrm{TF} = c_\textrm{TF} \, (1-p_b)/p_b$. Using that $c_\textrm{TF}=0.1\mu \mathrm M$  and $p_b=0.9$ gives $K_\textrm{op}^\textrm{TF} = 1.1\cdot 10^{-8}$~M.}.
This binding constant corresponds the binding energy $\epsilon_\textrm{op}^\textrm{TF} =-18.4~k_BT$  if using $\epsilon = k_BT \ln (K_\textrm{op}^\textrm{TF}/[1\mathrm M])$ \cite{sneppen2014models} ($k_BT$ is thermal energy), or $\epsilon_\textrm{op}^\textrm{TF}=-11.4$ kcal mol$^{-1}$ at room temperature ($k_BT=0.62\, \mathrm{kcal~mol}^{-1}$).  As discussed below,  $\epsilon_\textrm{op}^\textrm{TF}$ is almost four times smaller than the binding strength $\epsilon_\textrm{DNA}^\textrm{TF}$ associated with unspecific DNA binding.

As discussed above and explored in many theoretical models, TFs use a mixed strategy to find the designated target binding site  (e.g., \cite{von1989facilitated, elf2007probing, lomholt2009facilitated, hammar2012lac, benichou2011intermittent, klein2020skipping, mirny2009protein, hu2006proteins, hu2008dna, kolomeisky2011physics, hedstrom2023modelling} and many more). This strategy combines one-dimensional diffusional sliding along the DNA chain and three-dimensional jumps between DNA segments. By restricting a portion of the TF's search to several one-dimensional segments, the binding rate is effectively enhanced compared to a purely three-dimensional search. It improves because it is enough to find the correct DNA piece holding the target rather than diffusing directly into a nano-sized target from the bulk (a 10 bp target is only 2.7 nm long, 1 bp = 0.27 nm). Also, making three-dimensional jumps helps de-correlate the search, thus reducing repetitive visits to the same DNA sites, which characterizes a purely one-dimensional diffusive search. This combined process relies on a weak nonspecific affinity between TFs and DNA, much weaker than binding to an operator site ($K_\textrm{op}^\textrm{TF}/ K_\textrm{DNA}^\textrm{TF}\sim10^6$)%
\footnote{Using $\epsilon_\textrm{DNA}^\textrm{TF}\approx -4.8 k_BT$ for unspecific TF-DNA binding \cite{sneppen2014models} and $\epsilon_\textrm{op}^\textrm{TF} =-18.4 k_BT$, gives $K_\textrm{op}^\textrm{TF}/ K_\textrm{DNA}^\textrm{TF}= e^{(\epsilon_\textrm{DNA}^\textrm{TF}-\epsilon_\textrm{op}^\textrm{TF})/(k_BT)} \approx  8 \cdot 10^{5} \approx 10^{6}$.}.

But how strong should the non-specific interaction be to minimize the search? It should be long enough to ensure the target gets found if associated in the vicinity of the target but weak enough to allow TFs to quickly examine many segments and not get stuck. This problem is similar to an inspector looking for defective items (e.g., food) and wanting to know how much time should be invested in each one to find as many faulty items as possible in an enormous pile of candidate items. If going through them too quickly, there is a significant risk that faulty items go undetected. But if inspecting too slowly, there is not enough time to go through the pile. 

A similar issue arises in the so-called speed-stability paradox. Here, TFs scan the DNA sequence for specific binding motifs \cite{benichou2009searching, slutsky2004kinetics}. If the TF often pauses to check if it is on top of the target sequence, in other words, being very sensitive to the local base pair structure, it will find it with a probability of passing over the target. But, on the other hand, frequent pauses lead to a vanishingly small diffusion constant and long search times to distant targets.

This chapter explores how the TF search times (or binding frequencies) change with the length of the DNA stretches they examine before detaching. We call these segments antennae and consider a setup with millions of them, where only one harbors the designated target. We envision the target representing a gene regulatory sequence much shorter than the antenna itself. We open this chapter with a simple base case and calculate the flux through a single target lacking antenna. Next, we solve analytically a mathematical model capturing the rich interplay between unspecific DNA-TF binding of an extensive DNA segment pool and target-search times. Finally, we consider the search to a target on a single antenna and show that the search time has a non-trivial logarithmic correction with respect to its length.

\section{Binding rates to a single target}
\label{sec:sec2}
To define a reference case, imagine a target with radius $a$ floating in a cellular volume $V$ and a single searching protein. The protein randomly samples every target-sized 3D volume to find the target. If the trails are independent and the target volume fraction is $\phi = a^3/V$, the probability of finding the target after $n$ attempts is
\be
 \phi (1-\phi)^n,
\ee 
where the average number of trials is $\langle n \rangle$ is
\be
\langle n \rangle = \frac 1 \phi.
\ee
Next, assuming the protein explores each subvolume by diffusion, it spends approximately the time  $\tau = a^2/D_\textrm{3D}$ in each one, where $D_\textrm{3D}$ is the diffusion constant. The total search time  $T$ thus becomes $T= \langle n \rangle \times \tau$, or
\be
T  = \frac \tau \phi = \frac V {D_\textrm{3D} a}. 
\ee
By inverting this relationship, we obtain the corresponding stationary rate, or flux $J$
\be \label{eq:J_smol}
 J = D_\textrm{3D} a c,
\ee
where $c=1/V$ indicates a uniform protein concentration. Apart from a factor of $4\pi$, this equation is the famed Smoluchowski rate derived in numerous settings (e.g., in this issue\footnote{\textit{Target search on DNA - effect of coiling}, M. Lomholt.}).

Let's estimate $J$ from actual E. coli data.  Using that  $V = 1 (\mu\mathrm m)^3$, $c=100/V$ (100 proteins), $a=10$ nm (few base pairs wide target),  and $D_\textrm{3D} = 5 (\mu\mathrm m)^2 \mathrm{sec}^{-1}$ (nano-sized protein complex), gives the binding frequency $J=0.5~\mathrm{sec}^{-1}$ or $T=2$ seconds between rebinding events.

Next, we make the target-search problem more complex by letting the target be a small part of a DNA segment in a large pool of other segments to which proteins may bind and stay sequestered for significant durations.

\section{Target-binding rates in a pool of disconnected antennae}
\label{sec:facilitated}
As discussed in the Introduction, TFs search for target sequences by combining 3D diffusion and 1D sliding, where this mechanism builds on a non-specific DNA-protein attraction. However, there is a problem inherent with this setup. If proteins associate too strongly with DNA, they may spend significant time searching in the wrong places. On the other hand, if they happen to be close enough ("close" being within sliding distance or  "antenna length"), the proteins will detect it with almost certainty because they have a chance to diffuse over the target several times. Therefore, sifting through many DNA segments is critical to shortening the search time. But this requires lowering the unspecific binding energy.  And if too weak,  proteins start missing the target even if binding to the correct DNA segment because they dissociate too quickly. This poses an optimization problem that depends on the sliding or antenna length. Or, put differently, the optimal time spent searching each DNA piece. Below, we formulate a simple mathematical theory that allows us to explore this problem and calculate analytically the steady-state flux through a DNA target as a function of the unbinding rate $\koff$, which is proportional to the inverse of the unspecific binding time. We use E. coli as a primary example for our theory, where DNA permeates the entire cell volume. However, it applies equally to a cell nucleus in a mammalian cell that harbors the DNA (bacteria lack nuclei).

\begin{figure}
\includegraphics[width = \textwidth]{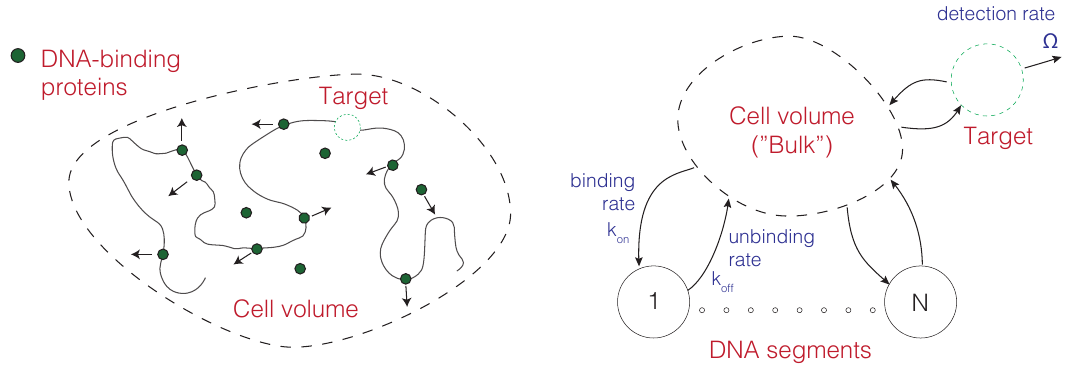}
\caption{
(left) Schematic cell interior where DNA (solid line) fills the cell volume (dashed). The figure also illustrates DNA-binding proteins (filled circles) diffusing in the cell volume and interacting with DNA to find a designated target sequence (open circle). 
(right) Effective DNA-target search model. The nodes $1,\ldots, N$ represent disconnected DNA segments (antennae) that are dynamically coupled via the cell volume ("Bulk"). The target node (dashed circle) is connected to the bulk like all other DNA nodes but has an additional detection or target-finding rate $\Omega$.
}
 \label{fig:effective-model}
\end{figure}

Consider the model sketched in Fig. \ref{fig:effective-model}. To the left, we depict the cell interior with DNA (solid line),  surrounding DNA-binding proteins (filled circles),  and a target (open circle). To the right, we illustrate an effective model suitable for analytical treatment. It shows DNA segments as unfilled circles or "nodes" ($i=1,\ldots, N$) that are dynamically coupled through the cell volume ("bulk"); The nodes are similar to the "DNA blobs" introduced in \cite{bauer2013vivo}. The arrows in Fig. \ref{fig:effective-model} highlight that proteins may bind and unbind DNA from and to the bulk.  We associate these arrows with on and off rates, denoted $\kon $ and $\koff$.  

The model omits proteins that translocate directly between nodes, so-called intersegmental transfer \cite{lomholt2009facilitated}. This  implies we treat the DNA segments as spatially scattered antennae that lack structural correlations related to DNA's 3D folding (e.g., as predicted by Hi-C experiments \cite{lieberman2009comprehensive, belton2012hi} or power-law distance decays in traditional polymer models). 

In addition to the DNA segments and the bulk, the figure shows the target as a separate node (dashed circle, far right). It has the same $\kon$ and $\koff$ as the other nodes but has also a target-detection rate $\Omega$.  Given $\koff$ and $\Omega$, the probability of finding the target sequence once a protein binds the correct node is
\be \label{eq:pb_target}
 \frac \Omega {\Omega + \koff}.
\ee
This equation shows that the detection probability drops to zero if $\koff$ is large. This suggests that $\koff$ be smaller to increase the detection probability. But if it becomes too low, the protein flux into the target reduces because most proteins stay bound to the other DNA nodes.

Below, we calculate how the flux depends on $\koff$ by recasting Fig. \ref{fig:effective-model} into a system of coupled equations for how the protein concentration in each node $c_i(t)$ changes over time $t$. Specifically, we are interested in the steady state flux $J_\infty = J( t\to \infty)$ through the target and how it changes with $\koff$.  We define  $J_\infty$ as
\be
J_\infty = \ctarget( t\to \infty) \Omega  \equiv \ctargetbar \Omega.
\ee

To calculate $\ctargetbar$, we first an write equation for $\ctarget(t)$,
\be \label{eq:ctarget}
 \frac{d \ctarget(t)}{dt}  = \kon \cbulk(t)  - (\koff + \Omega) \ctarget(t) ,
\ee
where $\cbulk(t)$ denotes the bulk protein concentration. Next, we write corresponding equations for each node $i=1,\ldots N$,  all following a similar structure:
\be
\begin{array}{ll}
 &\frac{d c_1}{dt} =  \kon\cbulk (t) - \koff c_1(t) \\
 &\ \ \ \vdots \\
 &\frac{d c_N}{dt} =  \kon \cbulk (t)  - \koff c_N(t).
\end{array}
\ee
By summing these equations, we may write a single equation for the protein concentration on DNA, $\cDNA(t)$, as
\be \label{eq:cDNA}
 \frac{d \cDNA(t)}{dt} =\frac d {dt}\big(c_1(t) + \ldots + c_N(t) \big) = \kon N \cbulk(t) -\koff  \cDNA(t).
\ee
In steady state, Eqs. \eqref{eq:cDNA} and \eqref{eq:ctarget} gives
\be \label{eq:cbar}
\ctargetbar = \frac{\kon}{\koff + \Omega} \cbulkbar, \ \ \
\cDNAbar = \frac{N \kon} {\koff} \cbulkbar.
\ee

Next, we relate $\ctargetbar$ to the cell's total protein concentration $c$. Because of mass conservation,  the proteins must either be sequestered on DNA, diffusing in the bulk, or attached to the target node. 
\be
 c = \ctarget(t) +\cDNA(t) + \cbulk (t) \approx \cDNA(t) + \cbulk (t).
\ee
We motivate the last step by noting that the target segment is a vanishingly small fraction compared to the rest of the genome ($\ll10^{-4}$ in bacteria). Using this equation and that $\cDNAbar$ from Eq. \eqref{eq:cbar}, we obtain
\be
 %
 \cbulkbar = \frac c {1 +\frac{N \kon}{\koff} },
\ee
and thus
\be
 \ctargetbar = \frac{\kon}{\koff + \Omega}  \frac c {1 +\frac{N \kon}{\koff} },
\ee
Using this expression to calculate the steady-state flux $J_\infty = \ctargetbar \Omega$, we obtain
\be 
 \label{eq:J_infinity}
 J_\infty  = \Omega \ctargetbar = \frac{1}{1 + \frac{\koff} \Omega } \frac{\kon c}{1 + \frac{N \kon}{\koff} }.
\ee

We note that each one of the factors in $J_\infty$ has intuitive interpretations. The first factor, $1/(1 + \koff/ \Omega)$, is the target node's reaction-limited contribution (or binding probability, Eq. \eqref{eq:pb_target}). When $\Omega$ is small, this factor drops to zero, indicating that the target requires a high revisiting frequency to reach short search times. The second factor, $ \kon c/(1 + N \kon/\koff )$ represents the reduced (Smoluchowski) diffusion flux ($\propto \kon c$) due to the proteins' being unspecifically bound elsewhere on DNA. 

Interestingly, the two factors in Eq. \eqref{eq:J_infinity} have an opposing dependence on $\koff$. This suggests that there exits an optimal $\kopti$ that maximizes the target flux $J_\infty^\mathrm{max} = J_\infty (\koff=\koff^*)$. We explore this optimum in the following section.

\subsection{Finding the optimal $\koff$ that maximizes target binding rates $J_\infty$}

\begin{figure}
\includegraphics[width = 0.5\textwidth]{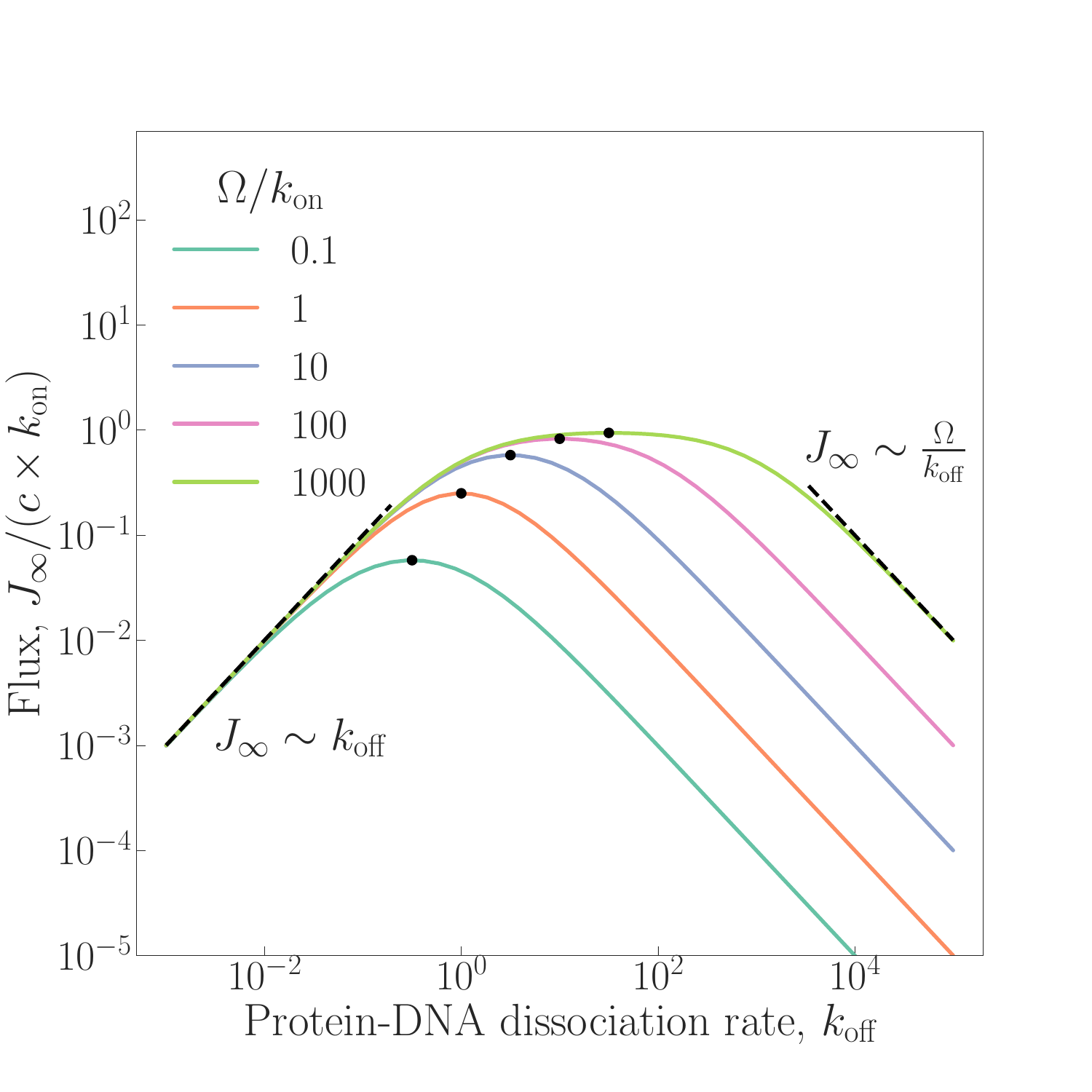}
\includegraphics[width = 0.5\textwidth]{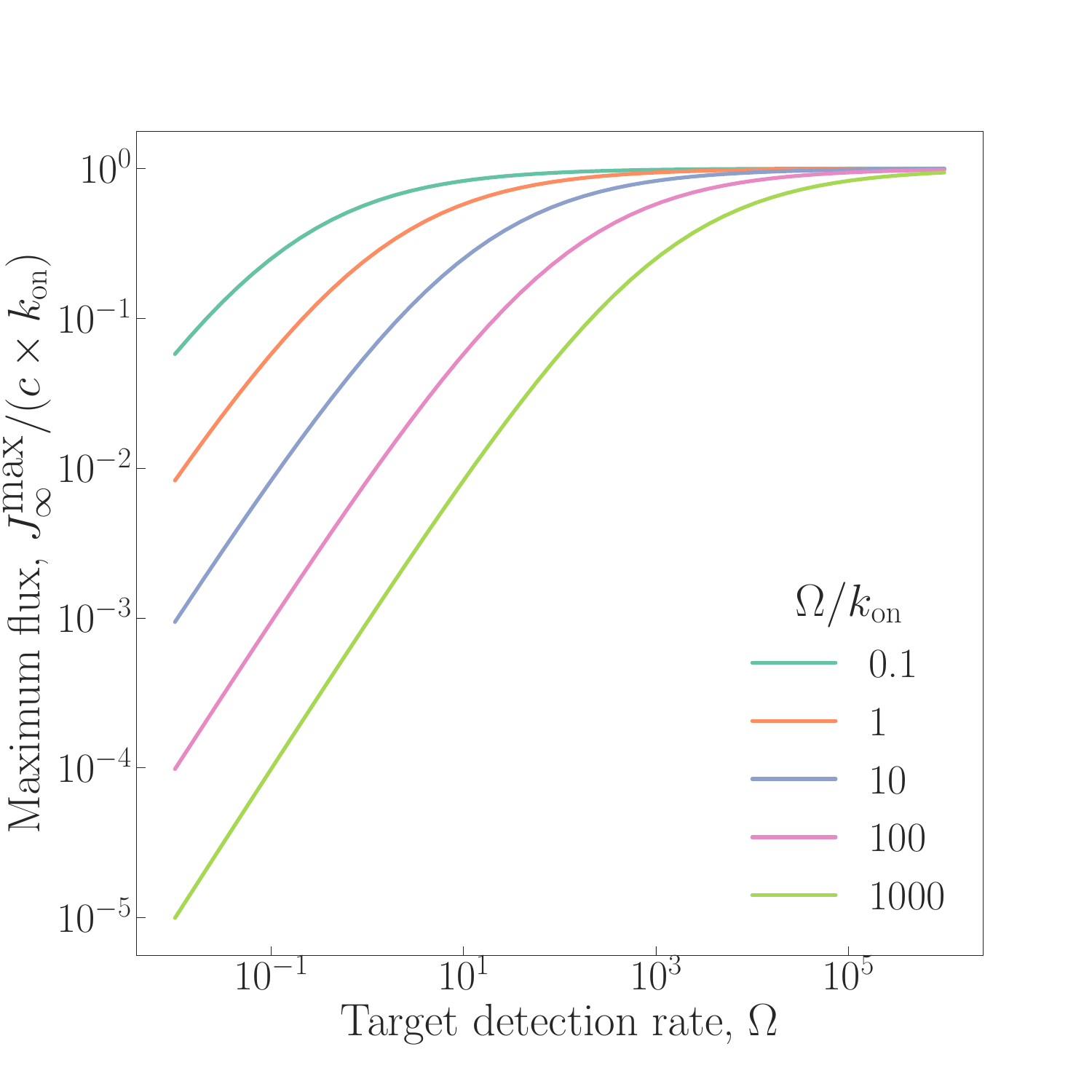}
\caption{%
(left) Target flux $J_\infty$ (Eq. \eqref{eq:J_infinity}) versus the rate $\koff$ of detaching from DNA for five detection rates $\Omega$. We plot asymptotic behaviors for small and large $\koff$ as dashed lines. For large $\koff$, we only plot the asymptote for $\Omega/\koff = 1000$ for clarity. The filled black circles indicate optimal points $(\kopti,  J_\infty^\mathrm{max})$ calculated from Eqs. \eqref{eq:koff_optimal} \& \eqref{eq:J_max}.
(right)  Maximal target flux versus target detection rate $\Omega$ for five on rates $\Omega/\kon$ (Eq. $\eqref{eq:J_max}$) . We note that the curves saturate for large $\Omega$, indicating that increasing the detection rate further does not lead to significantly higher flux.
}
\label{fig:flux-vs-koff}
\end{figure}

Before doing any analytical calculations, we plot the flux $J_\infty $ to see how it changes with the off-rate $\koff$ for a few detection rates $\Omega$ keeping  $N\kon$ ($=\konbar)$ fixed (Fig. \ref{fig:flux-vs-koff}, left). Besides the distinct maximum at $\kopti$, we note two regimes. These may be understood as follows. For small $\koff$, the flux increases linearly as $J_\infty \sim \koff$ (dashed line) in all cases. Here, most proteins remain sequestered on DNA for significant times, where the bulk concentration becomes low, $\cbulkbar \approx c \koff/\konbar$. This results in an $\Omega$-independent flux $(J_\infty \approx \kon \cbulkbar = (c/N)\, \koff$, thus limited by $\koff$ .
%

This contrasts with the other extreme when $\koff$ is large. Here, most of the proteins reside in the bulk, $\cbulkbar \approx c(1-\konbar/\koff)$, and the flux becomes $J_\infty \approx (\Omega/\koff)\,  \kon c$. In this case, proteins frequently find the correct segment but stay only for a short while, and the limiting rate is $\Omega$; the probability of finding the target is $\Omega/\koff \, ( 1 - {\cal O}((\Omega/\koff))$. We indicate the decay $\Omega/\koff$ as a dashed line.

Between these limiting cases resides an optimal $\kopti$  that maximizes the flux. To find this optimum, we calculate $\frac{d J_\infty }{d\koff} =0$ using Eq. \eqref{eq:J_infinity}, yielding
\be
 \frac{\Omega\kon c\, (\Omega\konbar - \koff^2)}{(\koff + \konbar)^2(\Omega+\koff)^2} = 0.
\ee

Solving this equation analytically gives (we omit the negative solutions)
\be 
  \label{eq:koff_optimal}
  \kopti =  \sqrt{\konbar \Omega} =\sqrt{N \kon \Omega}.
\ee
which gives the maximal rate $J_\infty^\mathrm{max} = J_\infty (\koff=\kopti)$ 
\be
 \label{eq:J_max}
 J_\infty^\mathrm{max} = \frac{\kon c}{\left(1 + \sqrt{\frac{N\kon} \Omega}\right)^2}.   
 \ee

We marked the optimal points $(\kopti, J_\infty^\mathrm{max}$) by symbols ($\bullet$) in Fig. \ref{fig:flux-vs-koff} (left). We also note that the plateau to the right optimal point becomes broader with increasing $\Omega$. We interpret this as if $\koff$ increases until the optimal point $\kopti$, increasing it more does not lead to a larger $J_\infty^\mathrm{max}$ because the flux is controlled by $\Omega$. But eventually, $\koff$ grows beyond $\Omega$, and the flux starts to decline as there is a significant chance of missing the target.

We also plotted how the maximal flux $J_\infty^\mathrm{max}$ changes with the target-finding rate $\Omega$ for varying on-rates $N\kon$ in Fig. \ref{fig:flux-vs-koff} (right). While $J_\infty^\mathrm{max}$ ranges over several orders of magnitude for small $\Omega$, it approaches the diffusion-limited Smoluchowski rate  $J_\infty^\mathrm{max}\simeq \kon c$ as the target becomes fully absorbing ($\Omega \to \infty$).

In biology, $\koff$ is inversely proportional to the protein-DNA residence time, making this parameter difficult to change. While it is conceivable that $\koff$ may change due to the binding of small molecules that alter the protein's binding affinity (e.g., allolactose binding to the Lac repressor), it is more likely that $\koff$ changes occur on evolutionary time scales.

\section{Target-binding rates to a single antenna with varying length}
The preceding discussion concerned diffusional flux through a target in the presence of many DNA segments that weakly bind the searching proteins. Here, we return to the single-antenna case and investigate how Smoluchowski's reaction rate through a target changes with DNA segment, or antenna, length $\lambda$. Specifically, we will study a case analytically and demonstrate that the flux has a non-trivial logarithmic correction $J_\lambda \propto \lambda/\ln \lambda$ when protein-antenna interactions are strong \cite{vasilyev2017smoluchowski}.

\begin{figure}
\includegraphics[width = \textwidth]{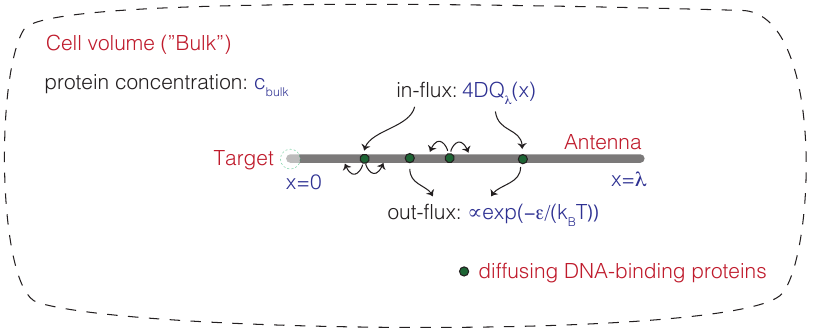}
\caption{
Schematic illustrating the single-antenna search problem. The antenna remains immersed (and immobile) in a large volume with constant protein concentration (filled circles). These proteins may bind to the antenna and diffuse while staying associated. After some time, they unbind to the bulk. We put the target at the antenna's left end ($x=0$). We assume the influx $4DQ_\lambda(x)$ and the outflux $\propto e^{-\epsilon/(k_BT)}$  are in steady-state.}
 \label{fig:single-antenna-model}
\end{figure}

To model this scenario, consider a volume with constant particle density and a suspended antenna with some length $\lambda$ (Fig. \ref{fig:single-antenna-model}). At one of its ends resides a target, which we represent as a small absorbing sphere. Like in the previous sections, we are interested in the steady-state conditions, where there is a balance between the diffusional flux from the volume onto the antenna and through the target. But unlike before, particles may bind to the target directly from the bulk without first associating with the antenna and then sliding (albeit this is a small contribution relative to the flux coming from the antenna).

Next, the antenna is associated with an interaction energy $\epsilon$ (or binding constant $K$) to particles diffusing in the surrounding volume. For now, we treat this interaction as very strong ($\epsilon\to \infty$ or $K\to 0$), implying that the particles never detach once associated with the antenna. However, at the end of this section, we relax this condition.

\subsection{Perfectly absorbing antenna}
To define the system's geometry, we use a 3D fcc lattice with coordinates $\mathbf r = (x, y, z)$, where $x, y, z$ are multiples of the lattice spacing $l_0$. The lattice spans symmetrically to infinity in all directions,  where we place the antenna in the origin and let it extend along the positive $x$-axis from $\mathbf r_0 = (0,0,0)$ to $\mathbf r_\lambda = (\lambda,0,0)$. To simplify notation,  we introduce linear coordinates $x$, where  $x=0$ and $x=\lambda$ are the antenna's two ends, and $\cAntenna = c(x,t\to \infty)$ is the effective linear particle density in steady-state. Note that the antenna has a finite thickness that equals the lattice spacing $l_0$, turning the antenna into a  rod ($l_0$ is also the target's diameter).

Far from the antenna, we assume the bulk concentration is constant $\cbulk$. We also assume the diffusion constant $D$  on the antenna and in the bulk are identical, that is $D=(l_0^2/6)\times k_\mathrm{jump}$ (admittedly, in actual cells, the diffusion constant may be orders of magnitude smaller). Below, we put the jump rate $k_\mathrm{jump}=1$ as we are interested in the flux-dependence on varying antenna lengths rather than diffusion rates. Also, we replaced $\lambda/l_0 \to \lambda$ to simplify notation ($l_0$ could represent one base pair).

To mathematically set up the problem, we formulate a one-dimensional diffusion equation for $\cAntenna$. In the strong-binding limit, we do not allow particles to escape the antenna back to the bulk. Therefore,  the flux into every inner antenna site from the four (out of six) adjacent bulk sites is $4\times D Q_\lambda(x)$, where  $Q_\lambda(x)$ is the density at those adjacent sites. This density interpolates the densities in the bulk $\cbulk$ and on the antenna $\cAntenna$. 

If approximating $x$ as a continuous variable, we model the particle density $\cAntenna$ as a diffusion equation with a linear line source term $4\times D Q_\lambda(x)$, representing the influx. That is,
\be \label{eq:diffusion_eq}
	D \frac{d^2\cAntenna}{dx^2} = -4D Q_\lambda(x), \ \ 0\leq x \leq \lambda
\ee
The absorbing condition at the target ($x=0)$ and flux boundary condition at the other end ($x=\lambda)$ are
\be \label{eq:bc1}
 \bar c(x=0) = 0,
\ee
\be \label{eq:bc2}
	D \left(\frac{d\cAntenna}{dx}\right)_{x=\lambda} = 5 D Q_\lambda(x=\lambda)
\ee
To get the last condition, we note that the flux at the antenna ends slightly differs from the flux into the inner nodes because the ends have five adjacent bulk nodes instead of four. This gives $5\times D Q_\lambda(x=\lambda)$. 

As a last step, we define the Smoluchowski rate through the target as
\be \label{eq:smol-rate-antenna}
	J_\lambda = - D  \left( \frac{d\cAntenna}{dx}\right)_{x=0},
\ee
which is the quantity of interest.

\begin{figure}
\includegraphics[width = \textwidth]{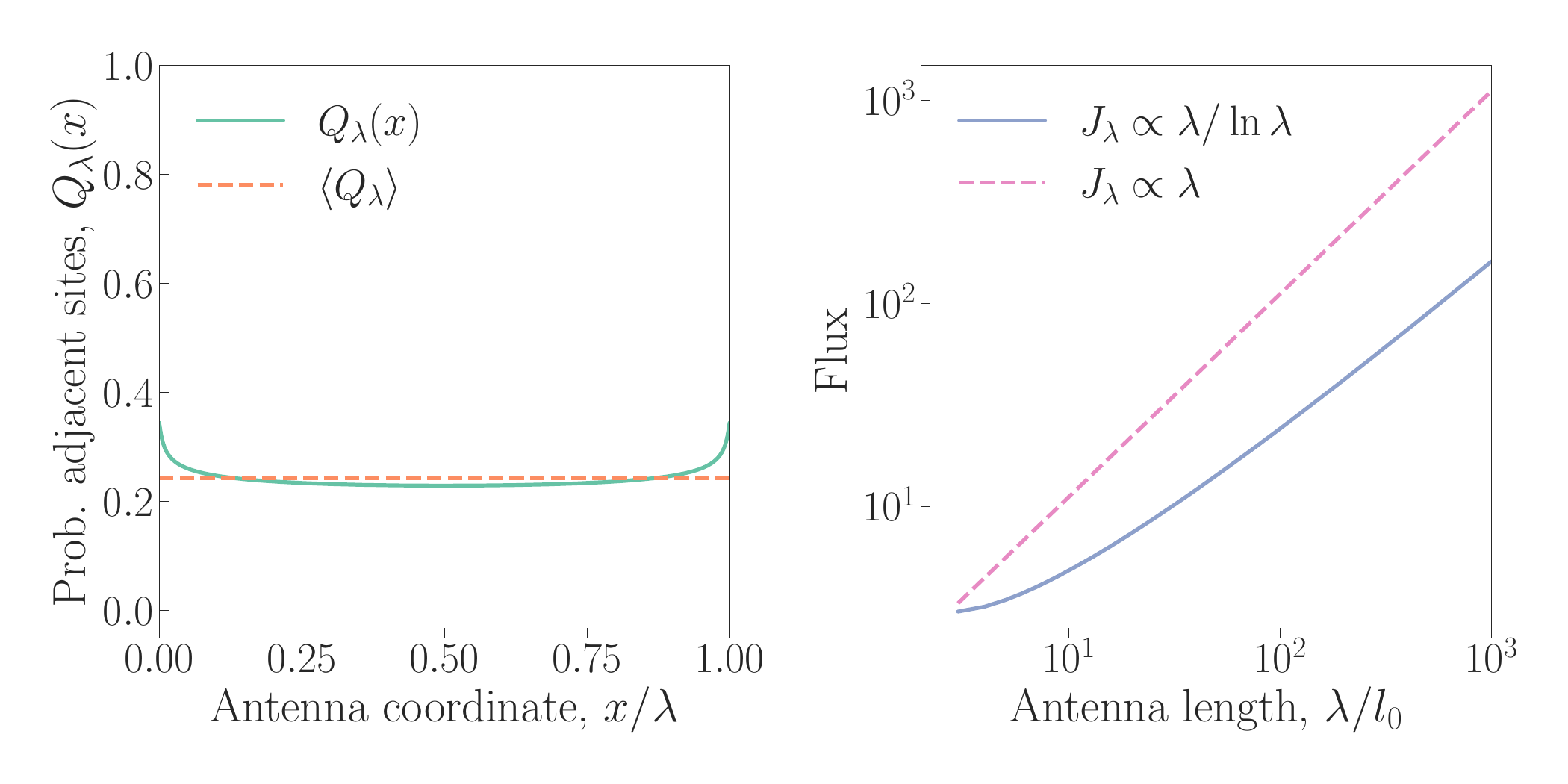}
\caption{
	(left) Particle density on sites adjacent to the antenna $Q_\lambda(x)$. The lines represent:  Eq. \eqref{eq:Q_vs_x} with $\cbulk=1$ (solid); antenna 
	average Eq. \eqref{eq:Q_average} (dashed).
	(right)  Flux through the target predicted from the theory Eq. \eqref{eq:J_lambda} (solid) and Smoluchowski's linear prediction (dashed).
	}
\label{fig:J_log}
\end{figure}

To calculate, $J_\lambda$, we must find  $Q_\lambda(x)$. As demonstrated in \cite{vasilyev2017smoluchowski},  this is a challenging and technical problem. However, it follows approximately
\be \label{eq:Q_vs_x}
 Q_\lambda(x) = \frac \cbulk {1+ \frac 1\pi (\textrm{arcsinh}(x) + \textrm{arcsinh}(\lambda - x))}.
\ee
Plotting this expression, we note that it is constant along most of the antenna, apart from the ends (Fig. \ref{fig:J_log}). Therefore, if the antenna is long, we may approximate  $Q_\lambda(x)$ by its average 
\be
  \langle Q_\lambda \rangle = \frac 1 \lambda \int_0^\lambda Q_\lambda(x)dx \simeq \frac A {1 + \frac 2\pi \ln \lambda }, \ \ \lambda \gg 1,
\ee
where $A \approx 1.06$. By using $Q_\lambda(x) \approx \langle Q_\lambda \rangle$ in Eq. \eqref{eq:diffusion_eq}, thus removing the explicit $x$-dependence in the source term, and using the boundary conditions in Eqs. \eqref{eq:bc1} \& \eqref{eq:bc2} (and $d^2\cAntenna/dx^2 = 4 \langle Q_\lambda \rangle$) to determine constants, we obtain
\be\label{eq:Q_average}
	\cAntenna = \langle Q_\lambda \rangle (-2x^2 + (4\lambda + 5)x).
\ee
Using this expression in Eq. \eqref{eq:smol-rate-antenna} to calculate the steady-state flux $J_\lambda$ through the target  gives 
\be \label{eq:J_lambda}
	J_\lambda \simeq -2\pi A D \frac{\lambda}{\ln \lambda}, \ \ \lambda \gg 1.
\ee	

This  result   highlights the logarithmic correction $[\ln(\lambda)]^{-1}$ in $J_\lambda$. As we portray in Fig. \ref{fig:J_log}, this correction substantially lowers the flux for long antennae relative to the linear increase $J\sim \lambda$ (Eq. \eqref{eq:J_smol}), representing Smoluchowski's prediction.

\subsection{Partially absorbing antenna}
To generalize these results to a partially absorbing antenna, we modify the diffusion equation Eq. \eqref{eq:diffusion_eq}. In particular, we change the right-hand side by adding an outflux term to the source
\be
	D \frac{d^2\cAntenna}{dx^2} = -4D \left(\langle Q_\lambda \rangle - e^{-\epsilon/(k_BT)} \cAntenna \right).
\ee
One of the boundary conditions in Eq. \eqref{eq:bc2} also changes. While the absorbing condition at $x=0$ stays intact, the flux condition at $x=\lambda$ modifies to
\be
	D \left(\frac{d\cAntenna}{dx}\right)_{x=\lambda} = 
	5 D \left( 
	    \langle Q_\lambda \rangle -   e^{-\epsilon/(k_BT)} \cAntenna 
	\right)_{x=\lambda}.
\ee
A relatively lengthy calculation gives that the leading $\lambda$ behaviour is \cite{vasilyev2017smoluchowski}
\be \label{eq:JJ_lambda}
 J_\lambda \simeq    J_0 \times \left[
 	1 +  \frac{2}{J_0} \langle Q_\lambda \rangle D e^{\epsilon/(2k_BT)} \tanh\left( \frac{2\lambda}{e^{\epsilon/(2k_BT)}}\right)
 \right],
\ee
where we introduced the shorthand $J_0$ denoting the standard Smoluchowski flux when the antenna length is zero, $J_{\lambda \to 1}$ (i.e., one single absorbing site in the 3D fcc lattice).

This expression allows us to explore $J_\lambda$ for varying $\epsilon$, keeping $\lambda $ large but fixed. First, we note that when $\epsilon\to\infty$, we recover our previous expression by expanding $\tanh \xi \approx \xi -\xi^3/3+\ldots$
\begin{equation}
 J_\lambda \simeq   J_0 \times  \frac{2}{J_0} \langle Q_\lambda \rangle D e^{\epsilon/(2k_BT)}  
  \left[ 
    \frac{2\lambda}{e^{\epsilon/(2k_BT)}} + \frac 1 3 \left(\frac{2\lambda}{e^{\epsilon/(2k_BT)}}\right)^3 + \ldots
 \right] 
\end{equation}
where
\begin{equation}
 |J_\lambda|   \sim   4\langle Q_\lambda \rangle D \lambda \sim  2\pi A D \frac{\lambda}{\ln \lambda}
\end{equation}

\begin{figure}
\includegraphics[width = \textwidth]{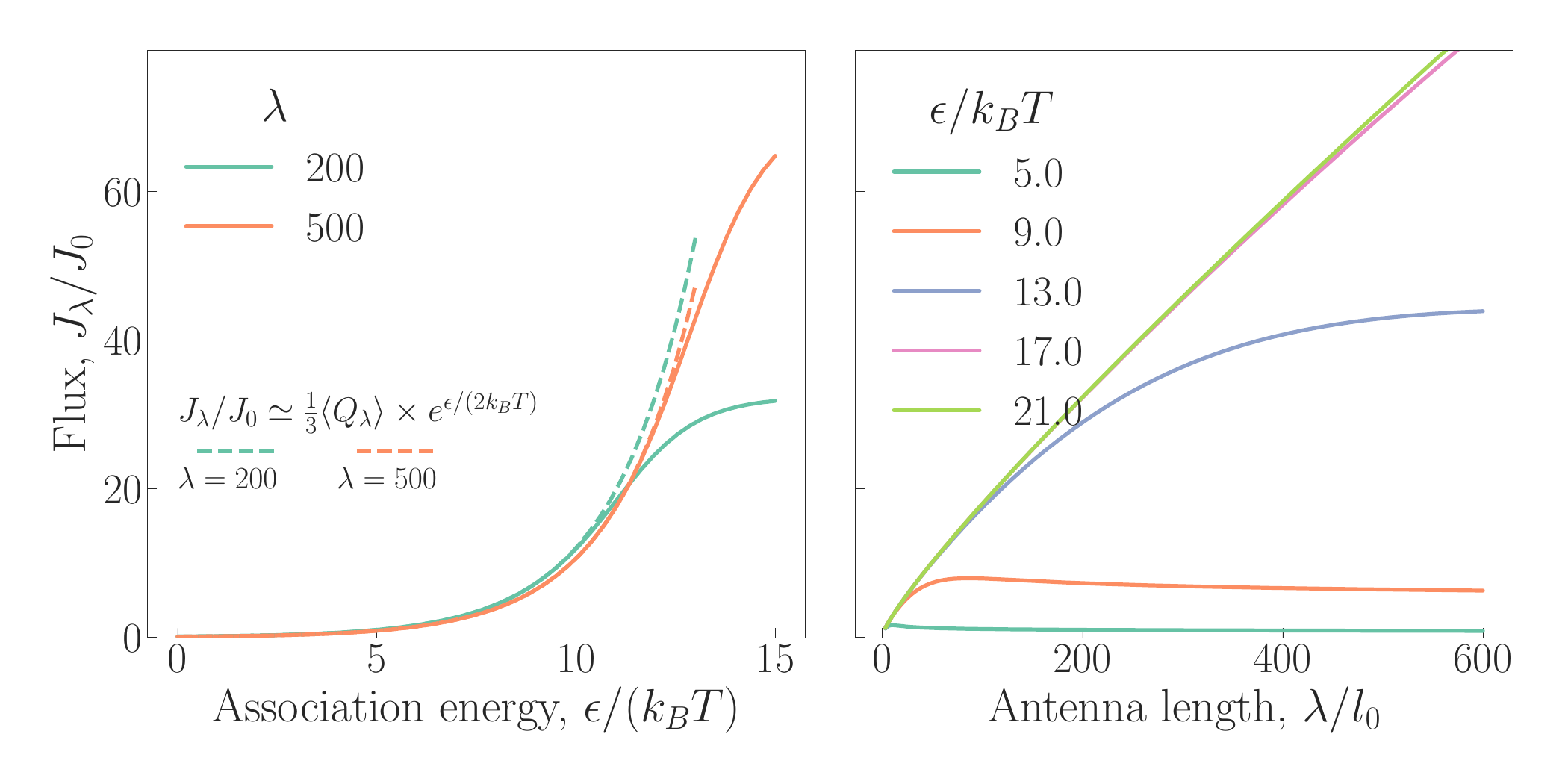}
\caption{
	(left) Target flux versus particle-antenna association energy for two different antenna lengths. The solid lines represent 
	Eq. \eqref{eq:JJ_lambda}, and the dashed lines highlight the approximation in Eq. \eqref{eq:JJ_approx}.
	(right) Target flux with increasing antenna lengths for five different particle-antenna association energies. As in the left panel, the solid lines represent 
	Eq. \eqref{eq:JJ_lambda}. Note that the lines cluster for increasing $\epsilon$, suggesting that there is a limiting $J_\lambda$ that 
	does not depend on $\epsilon$.
	}
\label{fig:J_epsilon}
\end{figure}

Furthermore, plotting Eq. \eqref{eq:JJ_lambda} reveals two noteworthy observations (Fig. \ref{fig:J_epsilon}). First, $J_\lambda$  increases rapidly for low to intermediate $\epsilon$.  To find this behavior analytically,  we note that the hyperbolic tangent approaches unity when  $2\lambda/(e^{\epsilon/(2k_BT)}) \gg 1$. Using this approximation gives an exponential dependence on $\epsilon$:
\be \label{eq:JJ_approx}
 |J_\lambda| \simeq 2 \langle Q_\lambda \rangle D e^{\epsilon/(2k_BT)} 
\ee
We plot this approximation as dashed lines alongside Eq. \eqref{eq:JJ_lambda}. We note that the match is better for longer antenna lengths.

Second, if instead plotting Eq. \eqref{eq:JJ_lambda} as a function of antenna length for a few fixed energies, we note that the $J_\lambda$-curves start to cluster and the flux becomes less and less dependent on $\epsilon$ (Fig. \ref{fig:J_epsilon}, right). This suggests that there is some characteristic $\epsilon_c$, above which increasing it more does not lead to a significantly larger flux, where the limiting maximal flux is 
\be
	J_\lambda^\mathrm{max} = \lim_{\epsilon \to \infty} |J_\lambda| \propto \lambda/\ln \lambda.
\ee

\section{Discussion and closing remarks}

The most common gene regulation mechanism is when a protein binds to a regulatory sequence to increase or decrease RNA transcription rates. However, these sequences are short relative to the genome length, so finding them poses a demanding search problem. This chapter presents two mathematical frameworks capturing different aspects of this problem. First, we studied the interplay between diffusional flux through a target where the searching proteins get sequestered on DNA out-of-reach from the target because of non-specific interactions. However, lowering these DNA-protein interactions increases the chance of missing the target, even if the proteins are close because of fleeting contacts. To make the model analytically solvable, we treated DNA as an ensemble of a disconnected antenna, thereby omitting structural correlations in 3D appearing naturally in polymer models or  Hi-C experiments \cite{belton2012hi, lieberman2009comprehensive}. We found that the optimal binding rate to be $(N\kon \Omega)^{1/2}$ ($N$ is the number of segments, $\kon$ is the protein association rate, and $\Omega$ is the target detection rate). Next, we studied the particle flux flows through a single antenna when changing its length. We found a non-trivial logarithmic correction to the linear behavior suggested by Smoluchowski's flux formula. 

The mathematical theories presented here rest on several assumptions. Below, we make a few remarks.

The model in Fig. \ref{fig:effective-model} divides DNA into $N$ disconnected segments, where we calculate the maximal flux $J_\infty^\mathrm{max}$ as a function of $\koff$ keeping $N$ fixed. Keeping $N$ fixed also means keeping the antenna length $\lambda$ fixed as $N\propto 1/\lambda$. However, under actual conditions, $\koff$  sets the proteins' DNA-residence time, making $\lambda$ and $\koff$ dependent variables. Using the 1D diffusion constant, we can estimate the antenna (or sliding) length as  $\lambda \propto (D_\mathrm{1D}/\koff)^{1/2}$. Thus, changing $\koff$ in principle modifies $N$ too. This observation leads to a more complicated mathematical theory that we leave as an open problem for interested readers to develop.

Next, we assumed that the antenna was straight. This assumption is realistic for transcription factors with sliding lengths shorter than DNA's persistence length $\approx$ 150 bp \cite{garcia2007biological}. However, several empirical studies report sliding lengths in the same order or longer (e.g., \cite{marklund2013transcription}). This situation suggests considering a coiled rather than a straight antenna. As demonstrated in a suite of papers \cite{hu2006proteins, hu2008dna}, it is possible to formulate a target search model for a coiled antenna following elegant but relatively intuitive arguments. Let's first consider the straight antenna. Smoluchowski's formula says the flux through a small sphere with radius $a$ is $J_\textrm{3D}\sim a$, where  $a\sim \lambda$ for a straight antenna. Next, like Section 4 in this chapter, we balance the bulk flux into the sphere with the particle flux a $J_\textrm{1D}$ onto the antenna inside the sphere. This flux sets the linear protein concentration on the antenna far from the target. Finally, knowing that the concentration vanishes at the absorbing target position, we may estimate the incoming linear flux $J$ using Fick's law. Now to the critical point in the argument: in steady-state, the fluxes on all scales must be in balance, thus
\be
 J_\textrm{3D} = J_\textrm{1D}=J.
\ee
This condition gives similar relationships as we derived in Section 2, from where it is possible to calculate the optimal $\lambda$ that minimizes the search. To generalize these arguments to a coiled antenna, it suffices to note that the sphere's radius $a$ is not linear in $\lambda$. Instead, because the antenna now is a coiled polymer, we associate $a$ with the polymer's radius of gyration, i.e., $a\sim \lambda^\nu$, where $\nu=1/2$ for a Gaussian coil. The rest of the analysis remains the same, and the optimal $\lambda$ now depends on $\nu$.

Also, we did not explore time-dependent transients associated with the particle flux or effects of initial conditions. But in E.coli, it seems "distance matters" \cite{pulkkinen2013distance}, where search times depend critically on where TFs are produced and how long they must diffuse to reach their designated targets. Some papers argue that this process drives genes to cluster spatially \cite{kolesov2007gene, di2013colocalization}.

Finally, we remark that the on-rate in Eq. \eqref{eq:cDNA}  easily generalizes to more complex DNA-node configurations than in Fig.  \ref{fig:effective-model}. If rewriting the total on-rate as 
\be
\konbar = \kon N = \kon e^{\ln N} \equiv  \kon e^{\Delta G/(k_BT)},
\ee
we may identify the free energy $\Delta G$ associated with the entropy of the node structures. For instance, the nodes may be wired into a complex  3D interaction network with scale-dependent community partitions \cite{bernenko2023mapping, holmgren2023mapping}. By including network configurations as free energy offers a framework to formulate effective target-search theories embracing space-filling fractal networks complementing previous efforts \cite{smrek2015facilitated, benichou2011facilitated, hedstrom2023modelling, nyberg2021modeling}.

Even if the DNA-target search problem is close to five decades old \cite{adam1968structural, riggs1970lac, von1989facilitated}, it still attracts new researchers, theoretical and empirical, across disciplines. We hope this chapter inspired future work exploring some unknown aspects of DNA-search processes or similar constrained search problems in other areas.


\end{document}